\documentstyle[epsf]{l-aa}     % LaTeX A&A  Standard Fonts
\begin{document}
   \thesaurus{11.19.2; 11.16.1; 11.09.4; 11.06.2; 04.01.1}
   \title{Extragalactic Database: VII}
 
   \subtitle{Reduction of astrophysical parameters}
 
\author{
G. Paturel \inst{1} \and
H. Andernach \inst{1} \and
L. Bottinelli \inst{2+3} \and
H. Di Nella \inst{1} \and
N. Durand \inst{2} \and
R. Garnier \inst{1} \and
L. Gouguenheim \inst{2+3} \and
P. Lanoix \inst{1}
M.C. Marthinet \inst{1} \and
C. Petit \inst{1} \and
J. Rousseau \inst{1} \and
G. Theureau \inst{2} \and
I. Vauglin \inst{1} 
}

    \offprints{G. Paturel}
 
   \institute{
              CRAL-Observatoire de Lyon, UMR 5574\\
              F69230 Saint-Genis Laval, FRANCE,\\
\and
              Observatoire de Paris-Meudon, URA 1757\\
              F92195 Meudon Principal Cedex\\
\and
              Universit\'e Paris-Sud\\
              F91405 Orsay, FRANCE\\
             }
 
   \date{Received September 10; accepted November 05, 1996}
%   \date{Paper in preparation - not for circulation}
 
   \maketitle
 
   \begin{abstract}
The Lyon-Meudon Extragalactic database (LEDA) gives a free access to
the main astrophysical parameters for more than 100,000 galaxies. The
most common names are compiled allowing users to recover quickly
any galaxy. All these measured astrophysical parameters are first
reduced to a common system according to well defined reduction
formulae leading to mean homogeneized parameters. 
Further, these parameters are also transformed into corrected parameters
from widely accepted models. For instance, raw 21-cm line widths 
are transformed into mean standard widths after
correction for instrumental effect and
then into maximum velocity rotation properly corrected for inclination
and non-circular velocity. This paper presents the reduction formulae
for each parameter: coordinates, morphological type and luminosity class,
diameter and axis ratio, apparent magnitude (UBV, IR, HI)
and colors, maximum velocity rotation and central velocity dispersion,
radial velocity, mean surface brightness, distance modulus and absolute
magnitude, and group membership. For each of these parameters intermediate 
quantities are given: galactic extinction, inclination, K-correction etc..

All these parameters are available from direct connexion to LEDA 
and distributed on  a standard CD-ROM (PGC-ROM 1996)  
by the Observatoire de Lyon via the CNRS.

   \end{abstract}
%================================================================== 

\section{Introduction}
This paper gives a detailed
description of the reduction  of astrophysical parameters available
through LEDA database for more than 100,000 galaxies. 
It is often required by users of LEDA who
need a reference where the description of parameters reduction is
given. Most of these reduction procedures were described in previous 
studies:
\begin{itemize}
\item Central velocity dispersion (Davoust et al., 1985)
\item Kinematical distance modulus (Bottinelli et al., 1986)
\item HI data (HI velocity, flux and 21-cm line width) (Bottinelli et al., 1990)
\item Diameters (Paturel et al. 1991)
\item Names of galaxies (Paturel et al., 1991a)
\item Group membership (Garcia, 1993)
\item Apparent magnitudes (Paturel et al. 1994)
\item Correction for inclination effect (Bottinelli et al., 1995)
\end{itemize}

Here, we will summarize these reduction and give reductions for some
additional parameters: morphological types and mean effective 
surface brightness.

Each forthcoming section will be devoted to a given class of parameter.

\section{General features}
In the following sections, each parameter will be designated in a unique
way by the word used in the LEDA query language. The correspondence
between this designation and the meaning is given in
Appendix B together with the variable name used in programs and the
format (FORTRAN convention).
Such a designation aims at avoiding subscript or exponant 
characters which can not be used in simple text printing or keyboard
entering.

The error on each parameter is now calculated using a method which gives
better description of the actual accuracy of the measurement. In previous
catalogs the standard error on an average parameter was simply
calculated from the sum of weights of each individual parameter
(the weight being the inverse square of individual standard error).
Now, the standard error is augmented
quadratically, by the external standard deviation between each parameter.
This estimate will be designated as the {\it actual uncertainty}. The precise
expression is explained in Appendix A. The main advantage of this new definition
is that a parameter with low actual uncertainty cannot result from 
discrepant individual measurements. So, this definition allows the user
to select undoubtly good data.

All parameters are available through LEDA 
\footnote{
 {\bf telnet} lmc.univ-lyon1.fr -- {\bf login: leda}\\
%or: telnet 134.214.4.7 -- login: leda
or: http://www-obs.univ-lyon1.fr/base/home\_base.fr.html  \\
}
or from a CD-ROM distributed by the Observatoire de Lyon via the CNRS.

\section{Name of galaxies}
We collected the most common names among 40. The different acronyms used
are listed in Table 1 with their abbreviation and the number of occurrences.
Some names designate several objects (e.g. UGC1 designates two galaxies,
PGC00177 and PGC00178). In such cases the name is given 
to both objects between parentheses.

\begin{table}
\label{Number of different acronyms used}
\begin{tabular}{lrl}
\hline
Acronym  &  N  &  Reference \\
\hline
PGC      &101258 & Paturel et al., 1989\\
MCG      & 30662 & Vorontsov-Velyaminov et al., 1962-1974,\\
CGCG     & 29825 & Zwicky et al., 1961-1968\\
ESO      & 17277 & Lauberts, 1982\\
UGC      & 13084 & Nilson, 1973\\
IRAS     & 11565 & IRAS, Point Source Catalogue, 1988\\
KUG      &  7942 & Takase \& Miyauchi-Isobe ,1984-1993\\
SAIT     &  7044 & Saito et al., 1990\\
NGC      &  6517 & Dreyer, 1888\\
DRCG     &  5725 & Dressler, 1980\\
IC       &  3509 & Dreyer 1895,1910\\
FGC      &  2573 & Karachentsev et al., 1993\\
VCC      &  2097 & Binggeli et al., 1985\\
FGCE     &  1881 & Karachentsev et al., 1993 \\
MARK     &  1514 & Markaryan et al., 1967-1981\\
KCPG     &  1206 & Karachentsev, 1987\\
ANON     &  1179 & de Vaucouleurs et al., 1976\\
FAIR     &  1185 & Fairall, 1977-1988\\
VV       &  1164 & Vorontsov-Velyaminov, 1977\\
nZW      &  2714 & Zwicky, 1971\\
KARA     &  1051 & Karachentseva, 1973\\
UM       &   652 & Kojoian et al., 1982\\
VIIIZW   &   645 & Zwicky et al. , 1975\\
ARAK     &   595 & Kojoian et al., 1981\\
KAZA     &   581 & Kazarian, 1979-1983\\
DCL      &   570 & Dickens et al., 1986\\
ARP      &   561 & Arp, 1966\\
HICK     &   464 & Hickson, 1993\\
UGCA     &   441 & Nilson, 1974\\
FCC      &   340 & Ferguson \& Sandage, 1990\\
FGCA     &   291 & Karachentsev et al., 1993 \\
SBS      &   284 & Markarian, 1983-1984\\
DDO      &   242 & Fisher \& Tully, 1975\\
WEIN     &   207 & Weinberger, 1980\\
TOLO     &   111 & Smith et al., 1976\\
RB       &    57 & Rood \& Baum, 1967\\
nSZW     &    58 & Rodgers et al., 1978\\
MESS     &    40 & Messier, 1781\\
POX      &    24 & Kunth et al., 1981\\
%NAMED    &    44 &\\
\hline
\end{tabular}
\caption{List of acronyms}
\end{table}

Since our first catalog of Principal Galaxies (PGC Paturel et al., 1989
, 1989a) we have added many new galaxies in LEDA database.
Each galaxy created in LEDA database receives a permanent LEDA number.
(with the acronym LEDA).
Note that LEDA number is identical to PGC number for running number
less than 73198. PGC numbers are sorted according to right ascension and
declination for epoch 2000. 

Many galaxies are known by their lexical name. These names are 
useful for some nearby large galaxies (Dwingeloo 1 and 2; Maffei 1 and 2
etc...). The equivalence of these names is given in Table 2.

%Table 1 acronyms
%Table 2 Lexical names

\begin{table}
\label{Lexical names}
\begin{tabular}{lll}
\hline
Lexical name    &Usual name    &PGC number\\
\hline
LMC             &ESO   56-115  &PGC 0017223  \\
SMC             &NGC     292   &PGC 0003085  \\
Maffei1         &UGCA     34   &PGC 0009892  \\
Maffei2         &UGCA     39   &PGC 0010217  \\
Circinus        &ESO   97- 13  &PGC 0050779  \\
%Holmberg1       &UGC    5139   &PGC 0027605  \\
%Holmberg2       &UGC    4305   &PGC 0023324  \\
%Holmberg3       &UGC    4841   &PGC 0026071  \\
%Holmberg4       &UGC    8837   &PGC 0049448  \\
%Holmberg5       &UGC    8658   &PGC 0048392  \\
%Holmberg6       &NGC    1325A  &PGC 0012754  \\
%Holmberg7       &UGC    7739   &PGC 0041861  \\
%Holmberg8       &UGC    8303   &PGC 0045927  \\
%Holmberg9       &UGC    5336   &PGC 0028757  \\
SextansA        &MCG -1-26- 30 &PGC 0029653  \\
SextansB        &UGC    5373   &PGC 0028913  \\
Carina          &ESO  206- 20A &PGC 0019441  \\
Draco           &UGC   10822   &PGC 0060095  \\
Fornax          &ESO  356-  4  &PGC 0010093  \\
%Leo1            &UGC    5470   &PGC 0029488  \\
%Leo2            &UGC    6253   &PGC 0034176  \\
Sculptor        &ESO  351- 30  &PGC 0003589  \\
UrsaMinor       &UGC    9749   &PGC 0054074  \\
%LGS1            &              &PGC 0001153  \\
%LGS2            &              &PGC 0001792  \\
%LGS3            &              &PGC 0003792  \\
%LGS4            &              &PGC 0001220  \\
%LGS5            &              &PGC 0001799  \\
Phoenix         &ESO  245-  7  &PGC 0006830  \\
LeoA            &UGC    5364   &PGC 0028868  \\
Pegasus         &UGC   12613   &PGC 0071538  \\
WLM             &MCG -3- 1- 15 &PGC 0000143  \\
Malin1          &              &PGC 0042102  \\
HydraA          &MCG -2-24-  7 &PGC 0026269  \\
CygnusA         &MCG  7-41-  3 &PGC 0063932  \\
HerculesA       &MCG  1-43-  6 &PGC 0059117  \\
Dwingeloo1      &              &LEDA0100170  \\
Dwingeloo2      &              &LEDA0101304  \\
\hline
\end{tabular}
\caption{Galaxies known by their lexical name}
\end{table}

\section{Coordinates}
Equatorial coordinates (Right ascension and Declination)
are given for two equinoxes 1950 (Besselian coordinates $al1950$, $de1950$) and
2000 (Julian coordinates $al2000$, $de2000$). Most of published coordinates 
are B1950 coordinates. Conversion to J2000 has been made according to the 
"Merits Standards" published in the U.S. Naval Observatory Circular. (1983).  
For computer use, coordinates are expressed as decimal values 
(hours to 0.00001 and degrees to 0.0001 for Right Ascension and Declination, 
respectively).
The standard deviation of coordinates is generally not known. Thus we
are using a flag $ipad$ to tell if the standard deviation is smaller than
10 arcsec or not. We collected systematically accurate coordinates in
literature (see Paturel et al. 1989). Recently, we added accurate coordinates
directly obtained from images stored in LEDA (Paturel et al., 1996)
and from COSMOS database (Rousseau et al., 1996).
Among the 100872 galaxies 69165 have accurate coordinates (69 percent).

\begin{figure}
\epsfxsize=8.5cm
\hbox{\epsfbox[40 400 350 650]{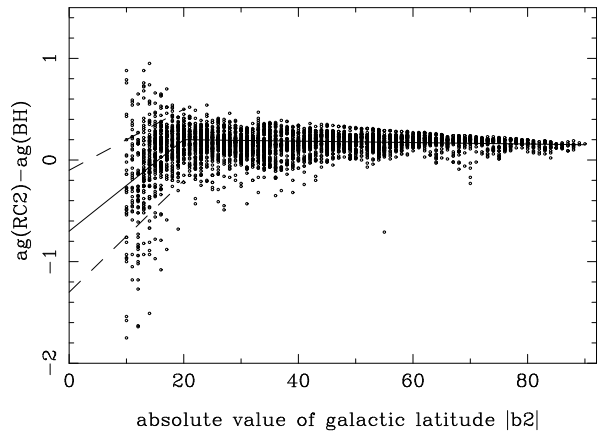}}
\caption{Difference of galactic extinction from RC2 and 
Burstein-Heiles vs. the absolute value of the galactic latitude.}
\label{ag}
\end{figure}

Galactic coordinates $l2$, $b2$ are calculated (in degrees to $0.01\deg$)
from $al1950$ and $de1950$ using the coordinates of the galactic pole 
$al1950$(pole)=12.81667 $de1950$(pole)=27.4000 and
the coordinates of the origin $al1950$(origin)=17.70667 $de1950$
(origin)=-28.9167
according to Blaauw et al. (1960). These galactic coordinates are
used to estimate the galactic extinction $ag$ converted to Burstein-Heiles
system (Burstein and Heiles, 1984) from the relationship given in the 
Second Reference Catalog (de Vaucouleurs, de Vaucouleurs and Corwin, 1976; 
p32, rel. 22; hereafter RC2).  In fact, for galactic latitude $b2 \geq 20 \deg$ the
difference between both systems is negligibly small (except for the zero
point difference of 0.20 mag due to the fact that Burstein-Heiles give
no absorption at the galactic pole).
In Fig.~\ref{ag} the difference between $ag$ from RC2 and from 
Burstein-Heiles is plotted {\it vs.} the galactic latitude.

The conversion from RC2- to BH-system is the following:\\
if $|b2|<20 \deg$:  
\begin{equation}
ag(BH) = ag(RC2) +  0.70 -0.045\ |b2| 
\end{equation}
and if $|b2| \geq 20 \deg$: 
\begin{equation}
ag(BH) = ag(RC2) +  0.001\ |b2|-0.22  \  
\end{equation}

The galactic extinction is higher than the one predicted by RC2 formula
for low galactic latitude ($b2 < 20 \deg$). The adopted galactic 
extinction is $ag=ag(BH)+0.20$, where $ag(BH)$ is calculated from Rel. 1 and 
Rel. 2.

Supergalactic coordinates $sgl$, $sgb$ are calculated (in degrees to $0.01\deg$)
from $l2$, $b2$ using the coordinates of the supergalactic pole
$l2$(pole)=47.37 deg $b2$(pole)=6.32 deg and the coordinates of the origin
$l2$(origin)=137.37 deg; $b2$(origin)=0 ~ deg according to 
de Vaucouleurs et al. (1976).

\section{Morphological type, luminosity class, and luminosity index}
Morphological types (E, SO, Sa ... Sm, Irr) have been entered in 
LEDA as an internal numerical code.
% according
%to the References Catalogs by de Vaucouleurs and co-workers. 
This code
will be treated as a continuous quantity. 

%It is defined in Table 3.
%
%\begin{table}
%\begin{tabular}{rlrl}
%\hline
%code &  morphological type & code &  morphological type \\
%\hline
%-5   &  E                  & 4    &  Sbc\\
%-3   &  E-SO               & 6    &  Sc\\
%-2   &  SO                 & 7    &  Scd\\
%0   &  SOa                & 8    &  Sd\\
%1   &  Sa                 & 9    &  Sm\\
%2   &  Sab                &10    &  Irr\\
%3   &  Sb                 & 5    &  S?  \\
%\hline
%\end{tabular}
%\caption{Input morphological type codes}
%\end{table}
 
Obviously, the definition of what is a given morphological type
is not the same for each astronomer. We thus adopted the RC3 type code system
as a reference one and converted to it all type codes
of others catalogs. A rms dispersion can be attached to each type code
for each reference allowing us to calculate a weighted mean morphological 
type code $t$ and its actual uncertainty $st$. 
In addition, some features have been coded in LEDA: ring, bar,
interaction (or multiplicity), and compactness. These features are
simply given as a flag R,B,M for the first three parameters and
C or D for the last one (C for compact and D for diffuse). 
Further, the numerical code $t$ and the features above are used
to produce a literal Hubble type $typ$ (e.g. SBa). The ranges
of definition are in Table 4.

\begin{table}
\begin{tabular}{rlrl}
\hline
range of $t$ &  $typ$ & range of $typ$ &  type \\
\hline
-5   $\leq t <$ -3.5  &  E                  & 3.5  $\leq t <$ 4.5   &  Sbc\\
-3.5 $\leq t <$ -2.5  &  E-SO               & 4.5  $\leq t <$ 6.5   &  Sc\\
-2.5 $\leq t <$ -1.5  &  SO                 & 6.5  $\leq t <$ 7.5   &  Scd\\
-1.5 $\leq t <$ 0.5  &  SOa                 & 7.5  $\leq t <$ 8.5   &  Sd\\
 0.5 $\leq t <$ 1.5  &  Sa                  & 8.5  $\leq t <$ 9.5   &  Sm\\
 1.5 $\leq t <$ 2.5  &  Sab                 & 9.5  $\leq t <$ 10    &  Irr\\
 2.5 $\leq t <$ 3.5  &  Sb                  \\
\hline
\end{tabular}
\caption{Output morphological type codes}
\end{table}

When the morphological type is uncertain ($st \geq 4.$) a rough type is
used with a question mark (e.g. E? or S?).

\begin{figure}
\epsfxsize=8.5cm
\hbox{\epsfbox[40 400 350 650]{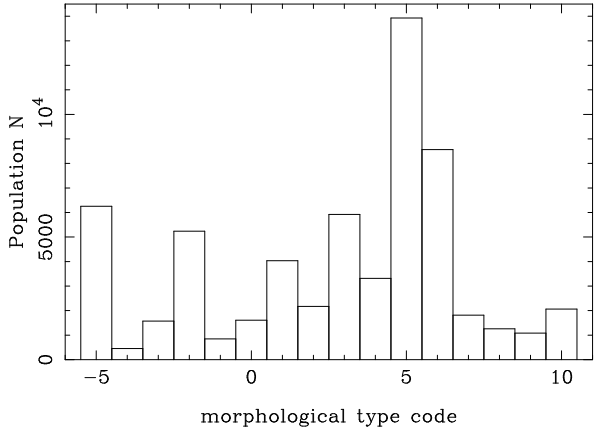}}
\caption{Histogram for morphological type codes.}
\label{htyp}
\end{figure}

The histogram of mean morphological type codes is given in Fig.~\ref{htyp}.
These type codes are available for 60130 galaxies.

The luminosity class introduced by Van den Bergh (1960) has been numerically
coded between 1 and 9 with the extension introduced by H.G. Corwin (SGC, 1985) 
between 9 and 11.
The codes are given in Table 5.
\begin{table}
\begin{tabular}{rlrl}
\hline
code &  original class  &  code  &  original class\\
\hline
 1   &  I         &  7     &  IV\\
 2   &  I-II      &  8     &  IV-V\\
 3   &  II        &  9     &  V \\
 4   &  II-III    & 10     &  V-VI \\
 5   &  III       & 11     &  VI \\
 6   &  III-IV     \\
\hline
\end{tabular}
\caption{Luminosity class codes}
\end{table}
This code is treated as a continuous parameter. We calculated 
the mean luminosity class $lc$ and the actual uncertainty $slc$ 
by giving the same weight to each reference.

Luminosity class codes and morphological type codes are used to
calculate the luminosity index $lambda$ according to de Vaucouleurs
et al. (1979). 

\begin{equation}
lambda = (lcc+t)/10
\end{equation}
where $lcc$ is the luminosity class corrected for inclination
according to the relation:

\begin{equation}
lcc=cl - c.logr25
\end{equation}

where $logr25$ is the axis ratio defined in section 6 and 
c is a parameter depending on the morphological type code $t$ ($c=2.0$
for $t\leq5$ and $c=2.0-0.9(t-5)$, otherwise).

\section{Diameter and axis ratio}
Many papers were devoted to the study of diameters and specially to the
reduction to the standard system defined by the isophote at the 
brightness of $25 B$-$mag\ arcsec^{-2}$. 
The conclusion of these studies was published by Paturel et al. (1991).

The diameters are expressed to $0.01'$ in log of $0.1'$ according to the convention 
of Second Reference Catalog (de Vaucouleurs et al., 1976). They are designated 
as $logd25$. For instance a diameter of $10'$ will be given as $logd25=2.00$.
Axis ratios are expressed in log of the ratio of the major axis to the minor 
axis. They are designated as $logr25$.

The main catalogs are reduced to the $D_{25}$-standard system using a relationship
\begin{equation}
    logd25 =  a.logD + b  
\end{equation}
\begin{equation}
    logr25 =  a'.logR 
\end{equation}
where D is the diameter and R is the ratio of the major axis 
to the minor axis in 
a given catalog. The constants $a$, $b$ and $a'$ are given in 
Paturel et al. (1991,
tables 1a and 1b) for the most common catalogs. 
Diameters and axis ratios extracted from LEDA  images or from COSMOS database 
were converted into the standard system using the same relationships but with 
different coefficients (Paturel et al., 1996; Garnier et al., 1996; 
Rousseau et al., 1996).

\begin{figure}
\epsfxsize=8.5cm
\hbox{\epsfbox[40 400 350 650]{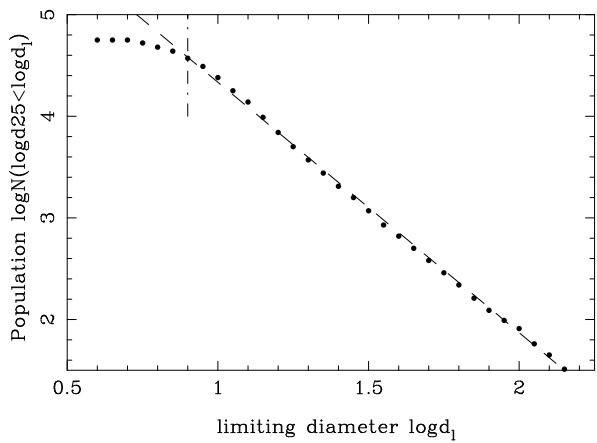}}
\caption{Completeness curve for $logd25$. The completeness is satisfied down
to the limit $logd_l=0.9$ (i.e. 0.8' in diameter).}
\label{cpl_d25}
\end{figure}

\begin{figure}
\epsfxsize=8.5cm
\hbox{\epsfbox[40 400 350 650]{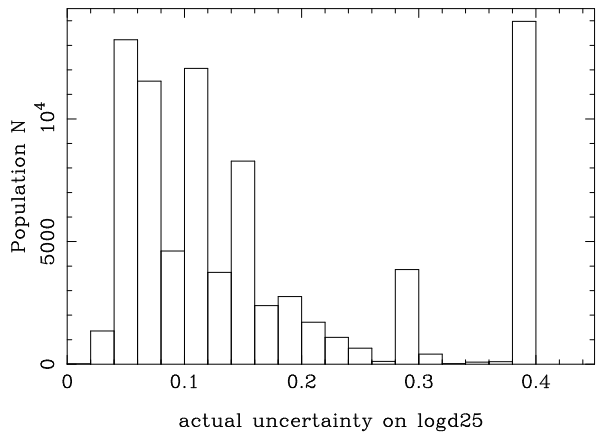}}
\caption{Histogram of actual uncertainty $slogd25$ on apparent diameter 
$logd25$.}
\label{hsid25}
\end{figure}

\begin{figure}
\epsfxsize=8.5cm
\hbox{\epsfbox[40 400 350 650]{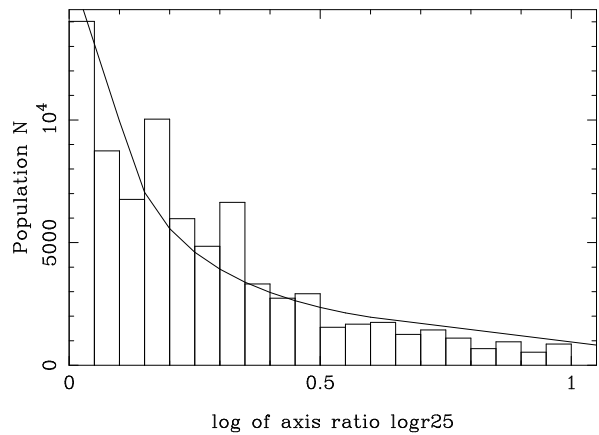}}
\caption{Histogram of log of axis ratio $logr25$. The solid
curve shows the distribution of $logr25$ for random orientation.}
\label{hr25}
\end{figure}

The completeness curve $logN$ {\it vs.} the limiting $logd25$ is shown in
Fig.~\ref{cpl_d25}.  The completeness is satisfied down
to the limit $logd_l=0.9$ (i.e. $0.8'$ in diameter).
Diameters $logd25$ are available for 82033 galaxies.
The histogram of actual uncertainty $slogd25$ on apparent diameter $logd25$
is given in Fig.~\ref{hsid25}. More than 13,000 galaxies have a diameter with
an actual uncertainty smaller than 0.05 (in $logd25$).
The distribution of logarithms of axis ratios is shown in Fig.~\ref{hr25}.
This distribution is close to the one expected if the orientation
of galaxies is randomly distributed.

\begin{figure}
\epsfxsize=8.5cm
\hbox{\epsfbox[40 400 350 650]{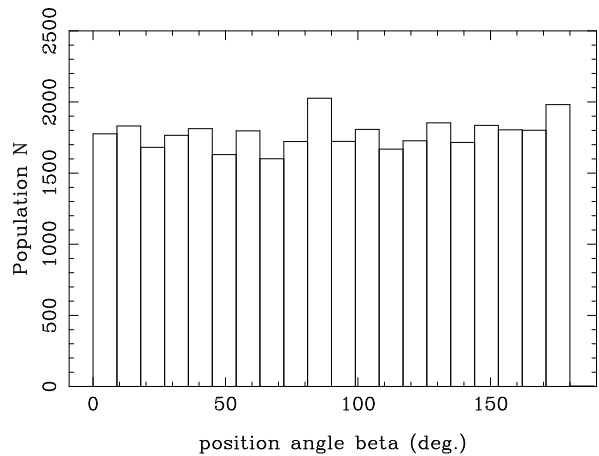}}
\caption{Distribution of major axis position angles.}
\label{hbeta}
\end{figure}

The position angle of the major axis is noted $pa$. It is 
counted from North towards East, between $0 \deg$ and $180 \deg$ and
is almost randomly distributed (Fig.~\ref{hbeta}). A small excess of 
galaxies appears at $pa=90\deg$ and $pa=180\deg$ which seems to be
an artifact. 

In RC2 apparent diameters were corrected for galactic extinction and inclination
effect according to Heidmann, Heidmann and de Vaucouleurs (1972abc). 
Recently, this question was revisited after the
result by Valentijn (1990, 1994) that galaxy disks are opaque. Our conclusion 
(Bottinelli et al. 1995) leads to the following correction:

\begin{equation}
    logdc =  logd25 - C.logr25 + ag.K_D
\end{equation}

where C=0.04, $ag$ is the galactic extinction (see section about coordinates) 
and $K_D$ is given by Fouqu\'e and
Paturel (1985) as $0.094$ for spiral galaxies and $0.081-0.016.t$ for
early type galaxies with morphological type code $t<0$.

\section{Apparent magnitude and colors}
The reduction of apparent B-magnitude to the RC3-system of magnitudes 
$B_T$ with photoelectric zero-point has been studied recently 
(Paturel et al., 1994). Apparent B-magnitudes reduced to the RC3-system
will be designated as $bt$.
Several effects were taken into account. The reduction of a given
magnitude $m$ to $bt$ is given by:

\begin{eqnarray}
\lefteqn{bt = a.m + b +} \nonumber  \\
 & &  c.(logr25-<logr25>) + d.(t-<t>) + \nonumber  \\
 & &  e.(logd25-<logd25>) + f.(de1950 -<de1950>)  \nonumber \\
\end{eqnarray}

where $a$, $b$, $c$, $d$, $e$, $f$, $<logr25>$, $<t>$, $<logd25>$, $<de1950>$ 
are constant values given in Paturel et al. (1994, table 6).
The mean $bt$ magnitude is calculated as a weighted mean
where the weight is derived for each source of magnitude
as the inverse square of the mean standard deviation. The final
actual uncertainty $sbt$ is derived from the total weight.

\begin{figure}
\epsfxsize=8.5cm
\hbox{\epsfbox[40 400 350 650]{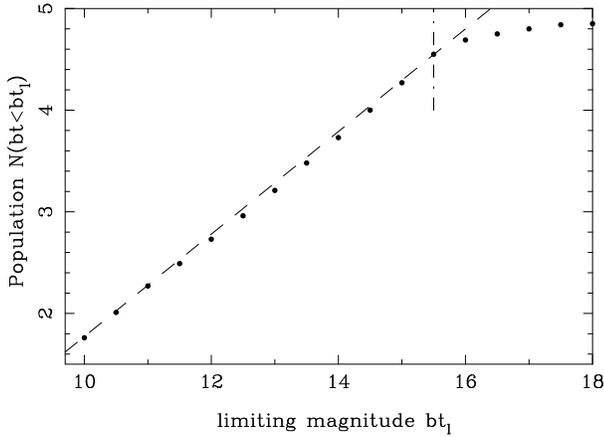}}
\caption{Completeness curve for apparent magnitude $bt$. 
The completeness is satisfied up
to the limit $bt=15.5$.}
\label{cpl_bt}
\end{figure}

\begin{figure}
\epsfxsize=8.5cm
\hbox{\epsfbox[40 400 350 650]{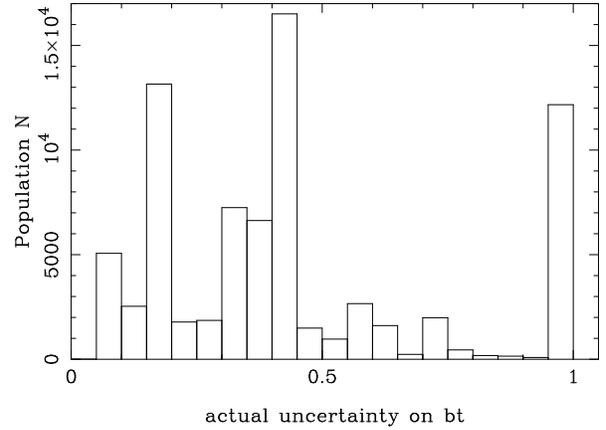}}
\caption{Histogram of actual uncertainty on apparent magnitudes $bt$.}
\label{hsibt}
\end{figure}

The cumulative completeness curve $logN$ {\it vs.} $bt$ is shown in
Fig.~\ref{cpl_bt}. The completeness in apparent magnitude is satisfied
up to $bt=15.5$. 
Apparent total magnitude $bt$ is available for 76760 galaxies.
The histogram of actual uncertainties $sbt$ on $bt$ is given in
Fig.~\ref{hsibt}. More than 7,000 galaxies have an actual uncertainty on $bt$
smaller than 0.02 mag.

Note that apparent diameter can be roughly converted into
a magnitude $m$ assuming that the mean surface brightness is constant for
all galaxies. The conversion can be made using the relationship (Di Nella
and Paturel, 1994):

\begin{equation}
m=20.0 - 5.logd25
\end{equation}

The standard deviation on $m$ is about $0.5 mag$. Using this relation,
it is possible to obtain an estimate of the apparent magnitude for 
93062 galaxies, 76760 magnitudes of which come from $bt$ and 16302 from
$logd25$. This magnitude $m$ will be used for drawing a more general
completeness curve (Fig.~\ref{cpl_m}).

Apparent $bt$ magnitudes are corrected for galactic extinction,
inclination and redshift effects according to the relation:

\begin{equation}
btc=bt-ag-ai-ak.v/10000
\end{equation}
where $ag$ is the galactic extinction in B (see section 4), expressed in 
magnitude, $ak$ is taken from de Vaucouleurs et al. (1976, RC2 p33, rel.25),
$v$  is the heliocentric velocity in $km.s^{-1}$ (section 9)
and $ai$ is given by Bottinelli et al. (1995) as:
 
\begin{equation}
ai= 2.5log(k + (1-k).R^{2C(1 + 0.2/K_D) - 1})    
\end{equation}

where, $k=l_{Bulge}/l_{Total}$ is taken from Simien and de Vaucouleurs (1986),
as a function of the morphological type code.
$K_D$ is taken from Fouqu\'e and Paturel (1985) as seen before (section 6), 
$C=0.04$ (Bottinelli et al., 1995) and $R=10^{alr25}$.
Note that this relation has been demonstrated for spiral galaxies only.
For early type galaxies ($t<0$) we assume $ai=0$, in agreement with
de Vaucouleurs et al. (1991).

Colors are given in the UBV system \footnote{some additional colors
$V-R$, $R-I$ are listed in LEDA but they do not represent a significant enough
sample to be included in the present version of the mean parameters}.
They are: 
total asymptotic colors $ubt$ for $(U-B)_T$ ; $bvt$ for $(B-V)_T$
and effective colors (i.e. colors within the effective aperture in which
half the total B-flux is emitted) $ube$ for $(U-B)_e$ ; $bve$ for 
$(B-V)_e$. 
Total asymptotic colors are corrected for galactic extinction,
inclination and redshift effects according to  RC3. The corrected
colors are $bvtc$ and $ubtc$ for $(B-V)_T$  and $(U-B)_T$ respectively.

\section{Maximum velocity rotation and central velocity dispersion}
We published compilations of HI-data in 1982 and 1990 (Bottinelli et al.,
1982; Bottinelli et al., 1990) but the data are regularly updated from
literature. The reduction of raw measurements is the same. The 21-cm line
widths are reduced to two standard levels (20\% and 50\% of the peak)
and to zero-velocity resolution using the following formula:
\begin{equation}
ws(l;r=0) = w(l',r) + (a.l'+b)r + c(l-l')
\end{equation}
where $ws(l,r=0)$ is the standard 21-cm line width at level $l=20$ or $l=50$,
while $w(l',r)$ is a raw measurement at a level $l'$ made with a velocity
resolution of $r\ km.s^{-1}$. The constants $a$, $b$ and $c$ are (Bottinelli
et al., 1990): $a=0.014, b= -0.83, c= -0.56$
 
The resulting standard 21-cm line widths $ws(l=20,r=0)$ and $ws(l=50,r=0)$ 
are corrected for systematic errors by intercomparison
reference by reference (program INTERCOMP, Bottinelli et al., 1982) leading to
standard widths $w20$ and $w50$ and their
actual uncertainties $sw20$ and $sw50$ respectively.

$w20$ and $w50$ are used to calculate the log of the maximum velocity 
rotation following the expression.
\begin{equation}
logvm= <log(wc)> -log(2sin(incl))
\end{equation}
where $incl$ is the inclination (in degrees) 
between the polar axis and the line of sight
calculated from the classical formula (Hubble 1926):
\begin{equation}
\sin^2(incl)= \frac {1-10^{-2.logr25}}{1-10^{-2.logro}}
\end{equation}
where $logro=0.43+0.053.t$,  if $-5\leq t\leq 7$ (or  $logro=0.38$ if $t>7$),
has been obtained from the most flattened galaxies.

$<log(wc)>$ is the weighted mean of the logarithm of 
the line widths $w20$ and $w50$ corrected for internal velocity dispersion. 
The adopted weight of level 20\% is twice the
weight of level 50\% because it is less sensitive to the definition
of the maximum and also because it corresponds to larger fraction of
the disk.
The correction for internal velocity dispersion is taken
according to Tully and Fouqu\'e (1985).
\begin{equation}
wc^2=w^2+wt^2(1-2e^{-w^2/wr^2})-2w.wt(1-e^{-w^2/wr^2})
\end{equation}
where $w$ is either $w20$ or $w50$ and 
$wt=2\sigma_z.k(l)$, assuming an isotropic distribution of the
non-circular motions $\sigma_z=12 km.s^{-1}$ and a nearly Gaussian velocity
distribution (i.e. $k(20)=1.96$ and $k(50)=1.13$).

Mean maximum velocity rotation $logvm$ is available for 6415 galaxies,
from 34,436 individual measurements $w20$ or $w50$.

\begin{figure}
\epsfxsize=8.5cm
\hbox{\epsfbox[40 400 350 650]{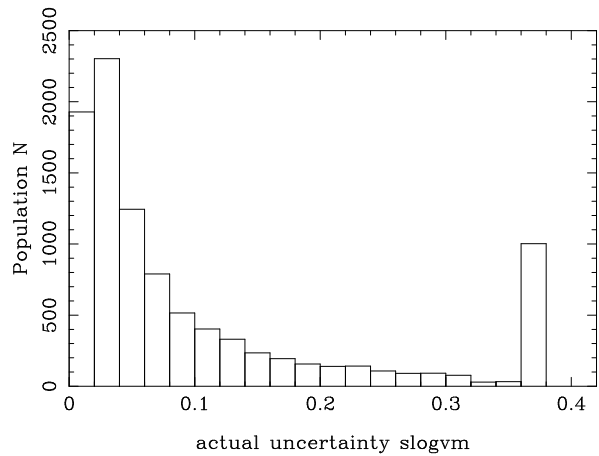}}
\caption{Histogram of the actual uncertainty on maximum velocity rotation 
$logvm$.} 
\label{hslogvm}
\end{figure}

The actual uncertainty on $logvm$ can be approximated  by
(For the detailed calculation see Bottinelli et al. 1983):
\begin{equation}
slogvm= 0.2\ \frac{sw^2} {w^2} + \frac {slogr25^2} {(10^{2 \: logr25}-1)^2}
\end{equation}
where $sw$ and $w$ are used for ($sw20$ or $sw50$) and ($w20$ or $w50$),
respectively.  
The histogram of $slogvm$ is presented in Fig.~\ref{hslogvm}.

\begin{figure}
\epsfxsize=8.5cm
\hbox{\epsfbox[40 400 350 650]{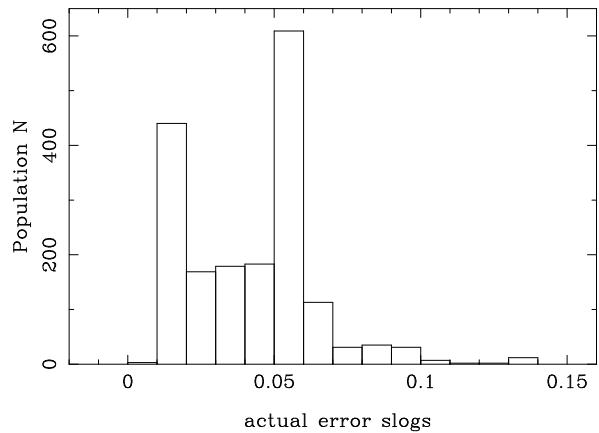}}
\caption{Histogram of the actual uncertainty on central velocity dispersion
$logs$.}
\label{hslogs}
\end{figure}

\vspace{1cm}
A preliminary compilation of central velocity dispersions $logs$  was
published in 1985 (Davoust et al.,1985) and included in our database.
This compilation has been regularly updated from literature (including
compilations made by Whitmore et al. 1985, McElroy 1995 and 
by Prugniel and Simien 1995).
Measurements from various references have been homogeneized using the
INTERCOMP program (Bottinelli et al., 1982). 
The mean central velocity dispersion $logs$  is available for 1816 galaxies
resulting from 3402 individual measurements. 
The actual uncertainty $slogs$ in log scale is shown in Fig.~\ref{hslogs}.

\begin{figure}
\epsfxsize=8.5cm
\hbox{\epsfbox[40 400 350 650]{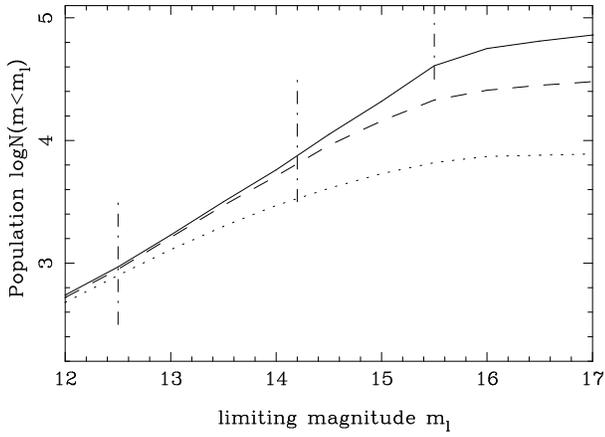}}
\caption{Completeness curve for $m$. The completeness is satisfied up
to the limit $m_l=15.5$ (solid line). 
This limit drop to $m \approx 14.2$ if we impose that the
radial velocity is known (dashed line), and to $m=12.0$ if we impose that 
either the maximum velocity rotation or the central velocity dispersion 
is known (dotted line).}
\label{cpl_m}
\end{figure}

In Fig.~\ref{cpl_m} we present the completeness of kinematical parameters
$logvm$ or $logs$ in comparison with the total completeness curve. 
The completeness is fulfilled up to about $m=12.0$ mag.

\section{Radial velocities}
Heliocentric radial velocities are obtained from 
optical or radio measurements $vopt$
or $vrad$, respectively.  The original optical compilation was made for 
the preparation of the 
RC3 catalog (Fouqu\'e et al., 1992). Velocities are corrected for systematic 
errors from the intercomparison reference by reference. The weight 
is deduced for each reference from this comparison. This allows the
calculation of the actual uncertainty $svopt$. 
Radio velocities
come essentially from 21-cm line measurements (plus some additional CO
measurements). The agreement between different authors is generally
excellent and there is no need of systematic correction. Mean error $svrad$
is given as a function of the velocity resolution (Bottinelli et al, 1990). 

From both $vopt$ and $vrad$ we calculate a weighted mean heliocentric velocity
$v$, the weights being the inverse squares of $svopt$ and $svrad$ respectively. 
The final weight leads to the actual uncertainty $sv$. When the discrepancy
between $vopt$ and $vrad$ is larger than $1000 km.s^{-1}$, we do not calculate
the mean heliocentric velocity $v$ but adopt instead 
the velocity having higher weight.  Radial velocity $v$ is available 
for 39667 galaxies.

From this mean heliocentric velocity $v$ we obtain four velocities defined
with different reference frames.
The velocity corrected to the galactic center $vgsr$ is obtained by a 
correction of the motion of the Sun with respect to the local standard 
of rest (LSR) and a correction of the LSR motion with respect to the
galactic center. The resulting correction is:

\begin{eqnarray}
\lefteqn{vgsr= v +232.sin(al2)cos(ab2)+} \nonumber \\
& & 9.cos(al2)cos(ab2)+ 7sin(ab2)
\end{eqnarray}

The velocity corrected to the centroid of the Local Group $vlg$ has been
adopted following Yahil et al (1977):

\begin{eqnarray}
\lefteqn{vlg = v +295.4sin(al2)cos(ab2)-} \nonumber \\
& & 79.1cos(al2)cos(ab2)- 37.6sin(ab2)
\end{eqnarray}

This correction replaces the classical IAU correction $300sin(al2)cos(al2)$.

The velocity corrected for infall of the Local Group towards Virgo 
is noted $vvir$. It is calculated as:

\begin{equation}
vvir=  vlg + 170.cos(\theta)
\end{equation}

where $170km.s^{-1}$ is the infall velocity of the Local Group according
to Sandage and Tammann (1990)
and where $\theta$ is the angular distance between
the observed direction $sgl$, $sgb$ in supergalactic coordinates and
the direction of the center of the Virgo cluster ($sglo = 104 \deg$, 
$sgbo = -2 \deg$).

\begin{eqnarray}
\lefteqn{cos(\theta)=sin(sgbo)sin(sgb)+} \nonumber \\
& & cos(sgbo)cos(sgb)cos(sglo-sgl)
\end{eqnarray}

Finally, the radial velocity is also expressed in the reference frame
of the Cosmic Background Radiation. This velocity is noted $v3k$. It is
calculated from the heliocentric velocity $v$ using the total solar
motion of $360 km.s^{-1}$ towards the direction defined by the 1950-
equatorial coordinates $al3k=11.25 h$ $de3k=-5.6\deg$ (Lubin and 
Villela, 1986). In 1997 this calculation should be replaced by the new
determination from COBE (Bennett et al. 1996). However, according to 
the rule defined at the end of the present paper (see the section 
"Acknowledgements"), the old definition will be used until the end of
1996:

\begin{equation}
v3k=  v + 360.cos(\theta 3k) 
\end{equation}

with $\theta 3k$ given by:

\begin{eqnarray}
\lefteqn{cos(\theta 3k)= sin(de3k)sin(de1950)+} \nonumber \\
& & cos(de3k)cos(de1950)cos(al3k-al1950)
\end{eqnarray}

\section{HI-line and IR fluxes} 
HI line flux (flux corresponding to the area under the 21-cm line
profile) and IRAS fluxes are treated separately from the classical
magnitudes (UBV) because they are obtained and corrected in a
completely different way.

The HI line flux is generally expressed in $Jy.km.s^{-1}$
converted in magnitude $m21$ according to the formula adopted in  RC3:

\begin{equation}
m21=-2.5 log(f) + 17.40
\end{equation}

where $f$ is the area of the 21-cm line profile expressed in 
$Jansky.km.s^{-1}$. This formula is equivalent to the one used
in RC3: $m21=-2.5log(fwm) + 16.6$, where $fwm$ is the flux in
$10^{-22}W.m^{-2}$. The standard error $sm21$ on $m21$ is given as a function
of the radiotelescope according to Bottinelli et al. (1990).

The HI line magnitude $m21$ has been corrected for self-absorption
effect following Heidmann et al. (1972):

\begin{equation}
m21c=m21 - 2.5log {\frac {\kappa /cos(incl)}{(1-\exp(\kappa /cos(incl))}}
\end{equation}

The adopted free parameter is $\kappa=0.031$. 
If inclination is higher than $89\deg$ the maximum correction is 
limited to $-0.82 mag$.

Both magnitudes $m21c$ and $btc$ are used to calculate a HI color index $hi$
initially defined by de Vaucouleurs et al. (1976): 

\begin{equation}
hi=m21c - btc
\end{equation}

\begin{figure}
\epsfxsize=8.5cm
\hbox{\epsfbox[40 400 350 650]{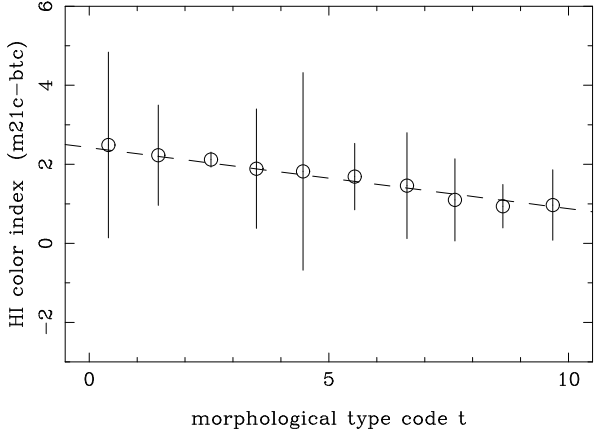}}
\caption{Mean HI color index $hi$ vesus morphological type code $t$.
}
\label{h1_m}
\end{figure}

This parameter is interesting as it is directly connected to the
hydrogen contents per unit of B-flux (see RC3, p51 Rel. 78).
The relation between $hi$ and morphological type code $t$ is presented
in Fig.~\ref{h1_m}. It shows a clear correlation which validates the
use of morphological type code as an observable parameter.

IRAS  fluxes at $60 \mu m$ and $100 \mu m$ are converted  
in the so-called far-infrared flux according to the relation:

\begin{equation}
mfir= -2.5log(2.58.f60 +f100) +14.75
\end{equation}

where $f60$ and $f100$ are IRAS fluxes at $60\mu m$ and $100\mu m$ 
expressed in Jansky. This relation is equivalent to the relation
given in RC3. The term 14.75 comes from the arbitrary zero-point of $20$
in RC3 (p.43, Rel. 49). The factor $2.58$ comes from IRAS {\it Point
Source Catalog} (1988).

\section{Mean surface brightness}
The parameter $brief$ (with its actual uncertainty $sbrief$) 
is the mean effective surface brightness, i.e. the mean surface 
brightness inside the effective aperture (the circular aperture 
enclosing one-half the total flux). This mean surface brightness 
is expressed in $B$-$mag.arcsec.^{-2}$.

Two equivalent measurements of $brief$ were derived: i) from the
apparent diameter $D_n$ enclosing a mean 
surface brightness of $20.75 B$-$mag.arcsec.^{-2}$ (Dressler et al. 1987), 
ii) from the mean surface brightness inside the effective isophote 
(elliptical isophote enclosing one-half the total flux) measured by 
Lauberts and Valentijn (1989; LV).

\begin{eqnarray}
\lefteqn{brief=-(1.87\pm 0.07) m'(D_n)+(60.13\pm1.36) }   \nonumber   \\
\lefteqn{\sigma = 0.49 }  \nonumber \\
\lefteqn{\rho = 0.74 \pm 0.02 }  \nonumber \\
\lefteqn{n=392 }  \nonumber \\
\end{eqnarray}

where $m'(D_n)=bt+5log(D_n)+4.38$, $\sigma$ is the standard deviation,
$\rho$ is the correlation coefficient and n the number of galaxies
used for the comparison. 

Similarly we have:

\begin{eqnarray}
\lefteqn{brief=(1.14\pm 0.02) m'(LV)+(3.05\pm0.46)+}    \nonumber  \\
& & \qquad \qquad \qquad \qquad \qquad \qquad \qquad (1.2\pm 0.1)logr25^2     \nonumber  \\
\lefteqn{\sigma = 0.49}   \nonumber \\
\lefteqn{\rho = 0.84 \pm 0.01}   \nonumber \\
\lefteqn{n=847}   \nonumber \\
\end{eqnarray}

where $m'(LV)$ is the {\it average blue central surface brightness
within half total B light} (noted $\mu_e(LV)$ in LV) , from Lauberts and 
Valentijn (1989).  The correction for inclination is in good agreement 
with the predicted one $1.35 \sim 1.26$ 
(see RC3 p50, Rel.71).  The final value $brief$ 
is calculated as the weighted mean of each determination.

Another estimate of the mean surface brightness is $bri25$, the mean surface 
brightness inside the isophote $25 B$-$mag.arcsec.^{-2}$.

This brightness  must be corrected for inclination
effect (Bottinelli et al., 1995). We then obtain the corrected mean
brightness $bri25$:

\begin{equation}
bri25=m'25 + 2.5 log(k.R^{-2C} + (1-k)R^{(0.4C/K_D)-1})
\end{equation}

where

\begin{equation}
m'25=bt+5 \ logd25 + 3.63
\end{equation}

Notations are those used for the calculation of $ai$ (section 7).

\section{Distance modulus and absolute magnitude}
The kinematical distance modulus $mucin$ can be derived from 
the heliocentric velocity properly corrected for the Local Group 
infall onto the Virgo cluster $vvir$
assuming a given Hubble constant $H_o=75 km.s^{-1}.Mpc^{-1}$. 
It must be noted that this distance does not include a correction for the infall
of individual galaxies onto Virgo. Such a distance could have been calculated
for instance using the model by Peebles (1976) as described in Bottinelli et
al. (1986). However, this model does not allow the calculation in the
direction of the Virgo center because of the third degree equation
(Eq. 2 in Bottinelli et al. 1986). Further, this model requires the
choice of a Virgo distance, of a Virgo mean radial velocity and of 
a velocity infall for the Local Group,
while the calculation of $mucin$ requires only the choice
of the velocity infall for the Local Group (we adopted 
$V_{infall}=170 km.s^{-1}$ ; see section 9). In the direction of Virgo cluster
center $mucin$ can be overestimated or underestimated depending on the
background or foreground position of the considered galaxy
with respect to Virgo center.

\begin{equation}
mucin= 5.log(vvir/75) +25
\end{equation}

$mucin$ is calculated only where $vvir >500 km.s^{-1}$, $mucin$ is
available for 39243 galaxies.
It is used to derive an estimate of the absolute magnitude $amabs$ in Blue band:

\begin{equation}
mabs=btc - mucin
\end{equation}

\section{Group membership}
The extragalactic database was used by Garcia (Garcia et al. 1993; Garcia 1993)
for a general study of group membership of all galaxies with a radial
velocity less or equal to 5500 $km.s^{-1}$ and an apparent magnitude brighter
than $bt=14$. Two automatic algorithms were used simultaneously 
(percolation and hierarchy clustering methods) for 
producing very robust groups. From the whole sample 485 groups were build.
They are identified by the acronym LGG (for Lyons Galaxy Group). 
For a galaxy, the LGG number $lgg$ gives the group to which the galaxy belongs.
The $lgg$ number is available for 2702 galaxies.

\acknowledgements{We are grateful to those who contributed to 
LEDA extragalactic database:
Becker M., Bravo H., Buta R.J., Corwin H.G., Davoust E.,
de Vaucouleurs A., de Vaucouleurs G., Fouqu\'e P., Garcia A.M.,
Kogoshvili N., Hallet N., Mamon G., Miyauchi-Isobe N., Odewahn S., 
Prugniel Ph., Simien F., Takase B., Turatto M. and many other people
who send some useful comments.

We want also to express our gratitude to some Institutions for their
financial support:
The "Minist\`ere de l'Enseignement Sup\'erieur et de la Recherche", 
The "Conseil R\'egional Rhone-Alpes" and the "Centre National de la Recherche
Scientifique". 

{\bf We would like to emphasize an important decision for
the future: we  will  maintain  the  same  reduction
procedures  for  a  full year (unless errors are found) in such a
way users can clearly  reference the data, for instance as LEDA1996.  
Any  changes will be announced in the LEDA news. The data of previous
years will be accessible on request}.
}

\vspace{0.5cm}
\appendix{{\bf Appendix A:} Calculation of the actual uncertainty.}

Let us assume that for a given galaxy we have $n$ measurements $x_i$
($i=1,n$) of a given parameter obtained from different references, 
each reference having a weight $w_i= 1/\sigma _i ^2$,
where $\sigma_{i}$ is the standard error of the i-th individual measurement. 
The actual uncertainty is calculated as:

\begin{equation}
\sigma_{a.u.}^2 = \sigma_{w}^2 + \sigma_{n}^2
\end{equation}

The first term $\sigma_{w}^2$ denotes the inverse of the total weight 
The total weight, $S_w$, is simply the sum of individual weights.
(i.e. $S_w= 1/\sigma_{w}^2= \sum 1/\sigma_{i}^2$).
This first term accounts for the accuracy of the reference of each 
individual measurement because the standard error of a given measurement
is assigned globaly from  e.g., the reference or the resolution etc... 
It is obvious that some individual measurements coming
from a good reference can be affected by a local problem (e.g., multiplicity
of the galaxy, star superimposed on the galaxy, bad seeing, misidentification
etc...). This fact will be taken into account by the second term.

The second term in the definition of the actual error is a measure of 
the consistency of the different measurements
building the mean measurement. It is calculated as the weighted standard 
deviation:
\begin{equation}
\sigma_{n}^2 = S_2/S_w  - (S_1/S_w)^2
\end{equation}
where :
$S_2 = \sum_{i} w_i x_i^2$, $S_1 = \sum_{i} w_i x_i$, $S_w = \sum_{i} w_i$.

The main advantage of the actual error is that it clearly shows any
internal uncertainty and any external discrepancy. 
 
\clearpage
\newpage
\onecolumn
\appendix{{\bf Appendix B:} LEDA's Astrophysical parameters }\\
\begin{verbatim}
--------------------------------------------------------------------------------
Parameter FORTRAN  FORTRAN columns  definition
name      name     format           
================================================================================
pgc       pgcleda  a11      1- 11   PGC or LEDA name (PGC = LEDA)           
ident     ident    a16     12- 27   1st name (NGC,IC,UGC,ESO...)           
ipad      ipadc    a1      28- 28   '*' for coordinates better than 10"
al1950    al1950   f10.5   29- 38   R.A. (B1950) (decimal hours)     
de1950    de1950   f10.5   39- 48   DEC. (B1950) (decimal degrees)    
al2000    al2000   f10.5   49- 58   R.A. (J2000) (decimal hours)          
de2000    de2000   f10.5   59- 68   DEC. (J2000) (decimal degrees)       
l2        al2      f10.3   69- 78   galactic longitude (degrees)        
b2        ab2      f10.3   79- 88   galactic latitude (degrees)        
sgl       sgl      f10.3   89- 98   supergalactic longitude (degrees) 
sgb       sgb      f10.3   99-108   supergalactic latitude (degrees)         
typ       typc     a5,1x  109-114   morph. type (e.g. 'E','Sab','SBa','SO')
morph     morphc   a4     115-118   'B' for Barred gal.   (see note below)
                                    'R' for Ring gal. 
                                    'M' for multiple gal. 
                                    'C' for compact, 'D' for diffuse
t         t        f10.3  119-128   morph. type code (-5 to 10)
st        st       f10.3  129-138   actual uncertainty on t           
lc        alc      f10.3  139-148   luminosity class (1 to 11) 
slc       slc      f10.3  149-158   actual uncertainty on lc          
logd25    alogd25  f10.3  159-168   log10 of isophotal diameter (d25 in 0.1')   
slogd25   slogd25  f10.3  169-178   actual uncertainty on logd25 
logr25    alogr25  f10.3  179-188   log10 of the axis ratio (major/minor axis)
slogr25   slogr25  f10.3  189-198   actual uncertainty on logr25           
pa        pa       f10.3  199-208   position angle (N->E) in degrees      
brief     brief    f10.3  209-218   effective surface brightness (mag.arcsec-2)
sbrief    sbrief   f10.3  219-228   actual uncertainty on brief                 
bt        bt       f10.3  229-238   total B-magnitude                         
sbt       sbt      f10.3  239-248   actual uncertainty on bt    
ubt       ubt      f10.3  249-258   (U-B)T                                      
bvt       bvt      f10.3  259-268   (B-V)T                                     
ube       ube      f10.3  269-278   (U-B)e                                    
bve       bve      f10.3  279-288   (B-V)e                                   
w20       w20      f10.3  289-298   21-cm line width at 20% of peak (in km/s)
sw20      sw20     f10.3  299-308   actual uncertainty on w20  
w50       w50      f10.3  309-318   21-cm line width at 50% of peak (in km/s) 
sw50      sw50     f10.3  319-328   actual uncertainty on w50          
logs      alogs    f10.3  329-338   log of the central velocity disp.(s in km/s)
slogs     slogs    f10.3  339-348   actual uncertainty on logs  
m21       am21     f10.3  349-358   HI-magnitude                                
sm21      sm21     f10.3  359-368   actual uncertainty on m21 
mfir      amfir    f10.3  369-378   far-infrared magnitude                    
vrad      vrad     f10.3  379-388   radio heliocentric radial velocity in km/s
svrad     svrad    f10.3  389-398   actual uncertainty on vrad       
vopt      vopt     f10.3  399-408   optical heliocentric radial velocity in km/s
svopt     svopt    f10.3  409-418   actual uncertainty on vopt 
v         v        f10.3  419-428   actual heliocentric radial velocity in km/s 
sv        sv       f10.3  429-438   actual uncertainty on v                    
--------------------------------------------------------------------------------
\end{verbatim}

\newpage
\onecolumn
\begin{verbatim}
--------------------------------------------------------------------------------
Parameter FORTRAN  FORTRAN columns  definition
name      name     format           
================================================================================
lgg       algg     f10.3  439-448   Lyon's galaxy group number             
ag        ag       f10.3  449-458   galactic extinction in B-mag 
ai        ai       f10.3  459-468   internal absorption (in B-mag)              
incl      aincl    f10.3  469-478   inclination
a21       a21      f10.3  479-488   HI self-absorption 
lambda    alambda  f10.3  489-498   luminosity-index 
logdc     alogdc   f10.3  499-508   log of the corrected diameter (dc in 0.1')
btc       btc      f10.3  509-518   corrected B-magnitude                    
ubtc      ubtc     f10.3  519-528   (U-B)o                                 
bvtc      bvtc     f10.3  529-538   (B-V)o                                
bri25     bri25    f10.3  539-548   mean surf. brightness within 25 m/"
logvm     alogvm   f10.3  549-558   log of max.circ. rot. vel.
slogvm    slogvm   f10.3  559-568   actual uncertainty on logvm
m21c      am21c    f10.3  569-578   corrected HI-magnitude 
hic       hic      f10.3  579-588   HI color index        
vlg       vlg      f10.3  589-598   radial vel. relative to the LG
vgsr      vgsr     f10.3  599-608   radial vel. relative to the GSR
vvir      vvir     f10.3  609-618   radial vel. corrected for Virgo infall
v3k       v3k      f10.3  619-628   radial vel. relative to the CBR
mucin     amucin   f10.3  629-638   kinematical distance modulus (H=75 km/s/Mpc)
mabs      amabs    f10.3  639-648   absolute B magnitude from mucin and mupar 
identi    identi   20a16  649-968   alternate names
--------------------------------------------------------------------------------
note: The parameter 'morph' can be read as 4 parameters (4a1 format)
      for Bar, Ring, Multiple and Compactness, respectively
\end{verbatim}

\end{document}